\documentclass[a4paper,11pt]{article}

\usepackage{amsfonts,latexsym,amsmath,amssymb,amsthm}
\usepackage[top=3cm,bottom=3cm,left=2.75cm,right=2.75cm,a4paper]{geometry}
\usepackage[colorlinks]{hyperref}
\hypersetup{
    colorlinks=true, 
    linkcolor={blue!50!black},
    citecolor={green!60!black},
    urlcolor= {red!80!black}
}
\usepackage{tikz,pgfplots,siunitx,multirow}
\usepackage{algorithmic,setspace,enumitem}
\usepackage{graphicx}
\usepackage[ruled,vlined]{algorithm2e}
\usepackage{empheq}
\usepackage[format=plain,justification=justified,singlelinecheck=off]{caption}
\usepackage[bottom]{footmisc} 
\usepgfplotslibrary{groupplots}
\usetikzlibrary{positioning,arrows,fit,patterns}
\usetikzlibrary{decorations.text}
\pgfplotsset{compat=newest}

\usepackage{subfig}
\DeclareCaptionLabelFormat{opening}{\textbf{#2)}} 
\usepackage{natbib} 

\makeatletter
\renewcommand*\env@matrix[1][\arraystretch]{%
  \edef\arraystretch{#1}%
  \hskip -\arraycolsep
  \let\@ifnextchar\new@ifnextchar
  \array{*\c@MaxMatrixCols c}}
\makeatother
\newcommand*\samethanks[1][\value{footnote}]{\footnotemark[#1]}

\newcommand{\myblue} {blue!80!white}

\newcommand{\myred}  {red}

\newcommand{\bx}          {\boldsymbol{x}}
\newcommand{\divergence}  {\nabla \cdot}
\newcommand{\tensS}       {\underline{\sigma}}
\newcommand{\tensE}       {\underline{\epsilon}}
\newcommand{\tensC}       {\underline{\underline{C}}}
\newcommand{\displacement}{\boldsymbol{u}}
\newcommand{\fsrc}        {\boldsymbol{f}}
\newcommand{\gsrc}        {\boldsymbol{g}}
\newcommand{\forward}     {\mathcal{F}}

\newcommand{\n}           {\boldsymbol{n}}

\newcommand{\misfit}      {\mathcal{J}}
\newcommand{\data}        {\boldsymbol{d}}
\newcommand{\obsU}        {\boldsymbol{d}_{\displacement}}
\newcommand{\obsS}        {\boldsymbol{d}_{\sigma}}
\newcommand{\obsSn}       {\boldsymbol{d}_{\sigma\cdot\n}}
\newcommand{\obsE}        {\boldsymbol{d}_{\epsilon}}
\newcommand{\simU}        {\forward_{\displacement}}
\newcommand{\simE}        {\forward_{\epsilon}}
\newcommand{\simS}        {\forward_{\sigma\cdot\n}}
\newcommand{\inddim}      {\mathsf{d}}
\newcommand{\bm}          {\boldsymbol{m}}

\newlength{\modelwidth}  \newlength{\modelRwidth}
\newlength{\modelheight} \newlength{\modelRheight}
\newlength{\plotwidth}   \newlength{\plotheight}
\newlength{\myhspace}
\newlength{\rbox}
\newcommand{\modelfile}{}
\newcommand{\mysubcaption}{}
\newcommand{\datafile}{}
\newcommand{\dataA}{}
\newcommand{\dataB}{}
\newcommand{\dataC}{}
\newcommand{\dataD}{}

\usepackage[capitalize,nameinlink]{cleveref}
\crefname{section}   {Section}   {Sections}
\crefname{subsection}{Subsection}{Subsections}
\Crefname{section}   {Section}   {Sections}
\Crefname{subsection}{Subsection}{Subsections}
\Crefname{figure}    {Figure}    {Figures}

\crefformat     {equation}{equation~\textup{#2#1#3}}
\crefrangeformat{equation}{equations~\textup{#3#1#4--#5#2#6}}
\crefmultiformat{equation}{equations~\textup{#2#1#3}}{ and \textup{#2#1#3}}
{, \textup{#2#1#3}}{, and \textup{#2#1#3}}
\crefrangemultiformat{equation}{equations~\textup{#3#1#4--#5#2#6}}%
{ and \textup{#3#1#4--#5#2#6}}{, \textup{#3#1#4--#5#2#6}}{, and \textup{#3#1#4--#5#2#6}}
\Crefformat     {equation}{#2Equation~\textup{#1}#3}
\Crefrangeformat{equation}{Equations~\textup{#3#1#4--#5#2#6}}
\Crefmultiformat{equation}{Equations~\textup{#2#1#3}}{ and \textup{#2#1#3}}
{, \textup{#2#1#3}}{, and \textup{#2#1#3}}
\Crefrangemultiformat{equation}{Equations~\textup{#3#1#4--#5#2#6}}%
{ and \textup{#3#1#4--#5#2#6}}{, \textup{#3#1#4--#5#2#6}}{, and \textup{#3#1#4--#5#2#6}}

\crefdefaultlabelformat{#2\textup{#1}#3}
\crefname{prop}{Proposition}{Propositions}
\Crefname{prop}{Proposition}{Propositions}
\crefname{defi}{Definition}{Definitions}
\Crefname{defi}{Definition}{Definitions}
\crefname{thm}{Theorem}{Theorems}
\Crefname{thm}{Theorem}{Theorems}
\crefname{rmk}{Remark}{Remarks}
\Crefname{rmk}{Remark}{Remarks}

\begin{document}

\title{Reciprocity-gap misfit functional for Distributed Acoustic Sensing, combining 
       data from passive and active sources}

\author{Florian Faucher\thanks{Faculty of Mathematics, University of Vienna, 
                                  Oskar-Morgenstern-Platz 1, A-1090 Vienna, Austria
                                  (\texttt{florian.faucher@univie.ac.at}).}
         \and
         Maarten V. de Hoop\thanks{Department of Computational and Applied Mathematics and 
                                     Department of Earth Science, Rice University, Houston TX 77005, USA.}
         \and 
         Otmar Scherzer\samethanks[1]\thanks{Johann Radon Institute for Computational and 
                Applied Mathematics (RICAM), Altenbergerstra{\ss}e 69 A-4040, Linz, Austria.}}

\date{}
\maketitle

\begin{abstract}
  Quantitative imaging of sub-surface Earth's properties in 
  elastic media is performed from Distributed Acoustic Sensing 
  data.
  A new misfit functional based upon the \emph{reciprocity-gap}
  is designed, taking cross-correlations of displacement and 
  strain, and these products further associate an observation 
  with a simulation.
  In comparison with other misfit functionals, this one has 
  the advantage to only require little a-priori information on the 
  exciting sources. 
  In particular, the misfit criterion enables the use of data from 
  regional earthquakes (teleseismic events can be included as well), 
  followed by exploration 
  data to perform a multi-resolution reconstruction.
  The data from regional earthquakes
  contain the low-frequency content
  which is missing in the exploration ones, allowing for the 
  recovery of the long spatial wavelength, even with very few sources.
  These data are used to build prior models for the subsequent 
  reconstruction from the higher-frequency exploration data. 
  This gives the elastic Full Reciprocity-gap Waveform 
  Inversion method, and we demonstrate its performance 
  with a pilot experiment for elastic isotropic reconstruction.
\end{abstract}
\section{Introduction}

Adjoint tomography and full waveform inversion in global and 
exploration seismology have, over the past two decades, dramatically 
improved our capabilities to estimate material properties and structure 
of Earth's interior, e.g.,~\cite{Tarantola1984,Gauthier1986,Pratt1998,
Ravaut2004,Tromp2005,Fichtner2006adjointa,Fichtner2006adjointb,
Tape2007,Liu2008,Fichtner2008,Luo2009,Virieux2009,Brossier2009,
Fichtner2010,Komatitsch2016} and \cite{Modrak2016}. 
In exploration seismology, the absence of relatively low frequencies 
in the data limits basic resolution \citep{Gauthier1986,Mora1987,LuoSchuster1991,Fichtner2008,Kroode2013}
unless some prior information is employed \citep{Bunks1995}. 
With the introduction of Distributed Acoustic Sensing (DAS) 
\citep{Mestayer2011,Cox2012,Daley2013,AjoFranklin2015}, 
a new opportunity has been provided to use (passive-source) global and
(active-source) exploration seismic data in an integrated manner, on
a regional scale, overcoming, in principle,
the resolution limitations. This is the subject of this paper.

In adjoint tomography and full waveform inversion, here
collectively referred to as FWI, the identification of 
Earth's parameters can be conducted via 
a minimization problem, 
where one optimizes a misfit criterion between observed 
data and simulations. 
It was first developed for the wave equation by 
\cite{Bamberger1977,Bamberger1979, Lailly1983} and \cite{Tarantola1984}, 
while the time-harmonic formulation was proposed by Pratt et al. 
(\citeyear{Pratt1996,Pratt1998, Pratt1999a}). 
The main difficulty comes from the local minima in the 
misfit functional, 
which are due to the inaccurate background velocity
information that causes phase shifts between observed and 
computed signals. This issue of cycle-skipping has
motivated several studies aiming at mitigating their 
occurrences, e.g., \cite{Gauthier1986,LuoSchuster1991,Bunks1995,Clement2001,Fichtner2008,Virieux2009} 
and \cite{Faucher2020Geo}. 
In particular, the frequency progression and multi-resolution 
strategy \citep{Bunks1995,Sirgue2004} have been widely applied. 
The use of different parametrizations of the inverse problem has
also been proposed by, e.g., \cite{Symes1991} and \cite{Clement2001}, 
at the cost of an increased computational complexity \citep{Faucher2020Geo}.
The presence of local minima in the misfit function 
and the convergence of iterative methods have also been studied 
theoretically through conditional, stability estimates 
\citep{Beretta2014,Beretta2017} in a multi-resolution 
framework \citep{deHoop2015,Beretta2016}.
However, if ``true'' Earth models are restricted to be piecewise smooth, 
there is a precise way to recover these through scattering control,
thus avoiding optimization \citep{Caday2019}.

While the misfit functional traditionally relies on the difference
between the full observations and the simulations, several
alternatives have been investigated. 
These include criteria based on the phase and 
envelope of the signal \citep{Fichtner2008}, the
cross-correlation of signals \citep{LuoSchuster1991, VanLeeuwen2010},
the $L1$ \citep{Brossier2010} and the optimal 
transport distance by \citep{Metivier2016,Yang2018}. 
In this paper, we propose a new functional based on the 
reciprocity theorem of time-correlation type \citep{deHoop2000}, 
using the displacement and the normal stress. 
Then, our misfit is equivalently written in terms of the strain, replacing the normal stress.
We coin it the \emph{reciprocity-gap} functional, following works 
on acoustic inverse scattering by \cite{KohnVogelius1985} and \cite{Colton2005}. 
\cite{Alessandrini2019} and \cite{Faucher2019FRgWI} establish theoretical
results for guaranteed recovery and provide applications in acoustic marine seismic.
Time correlation- and convolution-based functionals have
also been studied by, e.g., \cite{VanLeeuwen2010,Choi2011}
and \cite{Montagner2012}.
In addition to correlate the displacement and the stress, our 
misfit function also combines observation and simulation 
data.

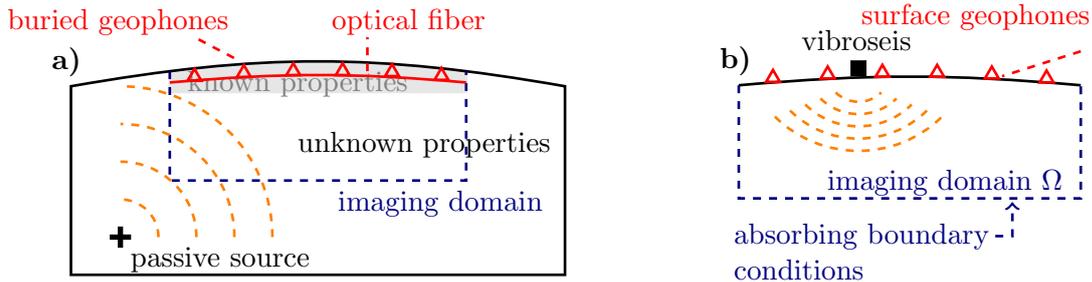
\begin{figure}[ht!] \centering
\subfloat{\label{fig:illustration:teleseismic} 
\begin{tikzpicture}[scale=1.]

  \pgfmathsetmacro{\wid}{6.5}
  \pgfmathsetmacro{\hei}{2.5}
  \coordinate (d1)   at (  0.,  0.);  
  \coordinate (d2)   at (\wid,  0.);
  \coordinate (d3)   at (\wid,\hei); 
  \coordinate (d4)   at (  0.,\hei); 
  \node[text width=0.5cm,anchor=south,black] at (0,\hei) {\textbf{a)}};

  \draw[line width=1,color=black]   (d1) to (d2) to (d3) to[out=170, in=10] (d4) to cycle;  

  \coordinate (r1)   at (0.20*\wid, 0.50*\hei);  
  \coordinate (r2)   at (0.80*\wid, 0.50*\hei);
  \coordinate (r3)   at (0.80*\wid, 1.08*\hei);
  \coordinate (r4)   at (0.20*\wid, 1.08*\hei); 
  \draw[line width=1,color=blue!50!black,dashed]   (r4) to (r1) to (r2) to (r3);  
  \node[text width=2.75cm,anchor=north east,color=blue!50!black,xshift=12mm] at (r2) {imaging domain};

  \coordinate (rlame1)   at (0.20*\wid,0.97*\hei); 
  \coordinate (rlame2)   at (0.80*\wid,0.97*\hei);
  \draw[line width=1,fill=black,opacity=0.10]  
    (rlame1) to (rlame2) to (r3) to[out=175, in=05] (r4) to  cycle;  
  \node[text width=3cm,anchor=south west,color=gray,yshift=-2mm,xshift=1mm] 
        at (rlame1) {known properties};
  \node[text width=4cm,anchor=south west,color=black,yshift=-1cm,xshift=1.55cm] 
        at (rlame1) {unknown properties};

  \coordinate (r3)   at (0.80*\wid, 1.02*\hei); 
  \coordinate (r4)   at (0.20*\wid, 1.02*\hei); 
  \draw[line width=1,color=red]   (r3) to[out=175, in=05] (r4);  
  \draw[mark=triangle,mark size=3,mark options={color=red,line width=1}] 
   plot[] coordinates{(0.25*\wid, 1.05*\hei)};
  \draw[mark=triangle,mark size=3,mark options={color=red,line width=1}] 
   plot[] coordinates{(0.35*\wid, 1.07*\hei)};
  \draw[mark=triangle,mark size=3,mark options={color=red,line width=1}] 
   plot[] coordinates{(0.45*\wid, 1.08*\hei)};
  \draw[mark=triangle,mark size=3,mark options={color=red,line width=1}] 
   plot[] coordinates{(0.55*\wid, 1.08*\hei)};
  \draw[mark=triangle,mark size=3,mark options={color=red,line width=1}] 
   plot[] coordinates{(0.65*\wid, 1.075*\hei)};
  \draw[mark=triangle,mark size=3,mark options={color=red,line width=1}] 
   plot[] coordinates{(0.75*\wid, 1.055*\hei)};

  \node[text width=3cm,anchor=south east,color=red,yshift=5mm,xshift=1cm] 
        at (r4) {buried geophones};
  \draw[line width=1,color=red,dashed]   (0.33*\wid, 1.15*\hei) to (0.25*\wid, 1.26*\hei);  
     
  \node[text width=3cm,anchor=south west,color=red,yshift=5mm,xshift=2cm] 
        at (r4) {optical fiber};
  \draw[line width=1,color=red,dashed]   (0.60*\wid, 1.08*\hei) to (0.60*\wid, 1.26*\hei); 

  \pgfmathsetmacro{\srcx}{0.1*\wid}
  \pgfmathsetmacro{\srcy}{0.2*\hei}  
  \draw[mark=+,mark size=4,mark options={color=black,line width=2}] plot[] 
        coordinates{(\srcx,\srcy)};
  \pgfmathsetmacro{\wsize}{0.5}   \pgfmathsetmacro{\wdepth}{0.5}
  \coordinate  (w1) at (\srcx+\wsize ,\srcy);
  \coordinate  (w2) at (\srcx        ,\srcy + \wdepth);
  \draw[orange,line width=1.,dashed](w1) to[out=90, in=0] (w2) ;
  \pgfmathsetmacro{\wsize}{1.00}   \pgfmathsetmacro{\wdepth}{1.00}
  \coordinate  (w1) at (\srcx+\wsize ,\srcy);
  \coordinate  (w2) at (\srcx        ,\srcy + \wdepth);
  \draw[orange,line width=1.,dashed](w1) to[out=90, in=0] (w2) ;
  \pgfmathsetmacro{\wsize}{1.50}   \pgfmathsetmacro{\wdepth}{1.50}
  \coordinate  (w1) at (\srcx+\wsize ,\srcy);
  \coordinate  (w2) at (\srcx        ,\srcy + \wdepth);
  \draw[orange,line width=1.,dashed](w1) to[out=90, in=0] (w2) ;
  \pgfmathsetmacro{\wsize}{2.00}   \pgfmathsetmacro{\wdepth}{2.00}
  \coordinate  (w1) at (\srcx+\wsize ,\srcy);
  \coordinate  (w2) at (\srcx        ,\srcy + \wdepth);
  \draw[orange,line width=1.,dashed](w1) to[out=90, in=0] (w2) ;

  \node[text width=3cm,anchor=north west,black] at (\srcx,\srcy) {passive source};

%
%
\end{tikzpicture}
} 
\hfill {\raisebox{-1mm}{
\subfloat{\label{fig:illustration:exploration} 
\begin{tikzpicture}[scale=1]

  \pgfmathsetmacro{\wid}{4.5} 
  \pgfmathsetmacro{\hei}{1.5} 
  \coordinate (r1)   at (  0.,  0.);  
  \coordinate (r2)   at (\wid,  0.);
  \coordinate (r3)   at (\wid,\hei); 
  \coordinate (r4)   at (  0.,\hei); 
  \coordinate (r5)   at (\wid,0.95*\hei); 
  \coordinate (r6)   at (  0.,0.95*\hei); 
  \coordinate (rlame1)   at (\wid,0.90*\hei); 
  \coordinate (rlame2)   at (  0.,0.90*\hei);
  \node[text width=0.5cm,anchor=south,black] at (0,\hei) {\textbf{b)}};
  \node[text width=3.2cm,anchor=south east,blue!50!black,yshift=-1mm] at (r2) 
       {imaging domain $\Omega$};
%
  \node[text width=3.50cm,anchor=north west,yshift=-2mm,blue!50!black,xshift=-2mm] 
        at (r1) {absorbing boundary conditions};
  \draw[line width=1,color=blue!50!black,dashed,->] 
  (0.75*\wid,-0.5) -| (0.8*\wid,-0.05);

  \draw[line width=1,color=blue!50!black,dashed]   (r4) to (r1) to (r2) to (r3);  
  \draw[line width=1]                    (r3) to[out=175, in=05] (r4);  
  \draw[mark=triangle,mark size=3,mark options={color=red,line width=1}] 
   plot[] coordinates{(0.10*\wid, 1.065*\hei)};
  \draw[mark=triangle,mark size=3,mark options={color=red,line width=1}] 
   plot[] coordinates{(0.26*\wid, 1.095*\hei)};
  \draw[mark=triangle,mark size=3,mark options={color=red,line width=1}] 
   plot[] coordinates{(0.42*\wid, 1.11*\hei)};
  \draw[mark=triangle,mark size=3,mark options={color=red,line width=1}] 
   plot[] coordinates{(0.58*\wid, 1.11*\hei)};
  \draw[mark=triangle,mark size=3,mark options={color=red,line width=1}] 
   plot[] coordinates{(0.74*\wid, 1.095*\hei)};
  \draw[mark=triangle,mark size=3,mark options={color=red,line width=1}] 
   plot[] coordinates{(0.90*\wid, 1.06*\hei)};

  \node[text width=3.75cm,anchor=south east,color=red,yshift=6mm,xshift=1cm] 
        at (r3) {surface geophones};
  \draw[line width=1,color=red,dashed]   (\wid, 1.3*\hei) to (0.77*\wid, 1.08*\hei);

  \pgfmathsetmacro{\srcx}{0.35*\wid}
  \pgfmathsetmacro{\srcy}{1.15*\hei}  
  \draw[mark=square*,mark size=2,mark options={color=black,line width=2}] plot[] 
        coordinates{(\srcx,\srcy)};
  \node[text width=1.5cm,anchor=south,black,yshift=1mm] at (\srcx,\srcy) {vibroseis};

  \pgfmathsetmacro{\wsize}{0.3}   \pgfmathsetmacro{\wdepth}{0.32}
  \coordinate  (w1) at (\srcx-\wsize ,\srcy - \wdepth);
  \coordinate  (w2) at (\srcx+\wsize ,\srcy - \wdepth);
  \draw[orange,line width=1.,dashed](w1) to[out=-45, in=225] (w2) ;
  \pgfmathsetmacro{\wsize}{0.5}   \pgfmathsetmacro{\wdepth}{0.4}
  \coordinate  (w1) at (\srcx-\wsize ,\srcy - \wdepth);
  \coordinate  (w2) at (\srcx+\wsize ,\srcy - \wdepth);
  \draw[orange,line width=1.,dashed](w1) to[out=-45, in=225] (w2) ;
  \pgfmathsetmacro{\wsize}{0.70}   \pgfmathsetmacro{\wdepth}{0.48}
  \coordinate  (w1) at (\srcx-\wsize ,\srcy - \wdepth);
  \coordinate  (w2) at (\srcx+\wsize ,\srcy - \wdepth);
  \draw[orange,line width=1.,dashed](w1) to[out=-45, in=225] (w2) ;
  \pgfmathsetmacro{\wsize}{0.90}   \pgfmathsetmacro{\wdepth}{0.56}
  \coordinate  (w1) at (\srcx-\wsize ,\srcy - \wdepth);
  \coordinate  (w2) at (\srcx+\wsize ,\srcy - \wdepth);
  \draw[orange,line width=1.,dashed](w1) to[out=-45, in=225] (w2) ;
  \pgfmathsetmacro{\wsize}{1.10}   \pgfmathsetmacro{\wdepth}{0.64}
  \coordinate  (w1) at (\srcx-\wsize ,\srcy - \wdepth);
  \coordinate  (w2) at (\srcx+\wsize ,\srcy - \wdepth);
  \draw[orange,line width=1.,dashed](w1) to[out=-45, in=225] (w2) ;
  
\end{tikzpicture}
}}} 
\caption{Illustration of the configuration where
         \protect\subref{fig:illustration:teleseismic}
         the source of the regional earthquake is 
         allowed to lie outside of the 
         \protect\subref{fig:illustration:exploration}
         computational domain in exploration.
         In order to avoid the free-surface condition 
         where the normal stress is zero, the 
         devices are considered slightly buried, and 
         we assume the knowledge of the physical properties 
         in the near surface area (in grey) for FRgWI.         
         }
\label{fig:illustration}
\end{figure}

The key feature of our underlying Full Reciprocity-gap 
Waveform Inversion (FRgWI) method is that the misfit functional 
we introduce does not require knowledge 
(location and characteristics) of the passive and active sources
(that is, the sources from which the measurements are extracted). 
We exploit these properties in integrating 
regional earthquakes 
and exploration data from a
natural multi-resolution perspective, that we illustrate in \cref{fig:illustration}. 
Regional earthquakes (and teleseismic) 
data contain the very low frequencies that are missing, 
and needed, in the exploration setting.
Furthermore, we shall see that a small number of sources is sufficient  to
extract the long wavelength profile of the media.
In seismic inversion using data from teleseismic events 
(distant earthquakes), it is assumed that the (teleseismic) 
source is sufficiently distant from the area of interest 
so that the incoming waves that illuminate this area can 
be considered as plane waves, 
cf.~\cite{BostockShraggeRondeany2001a} and \cite{Tromp2019Review}.
Under such an assumption, the characterization of the 
distant source is not needed for imaging and, for instance, 
reconstructions using teleseismic data is carried out by 
\cite{BostockShraggeRondeany2001b} and \cite{BostockShraggeRondeany2001c}
assuming the Born approximation, and by \cite{Beller2018a,Beller2018b} 
using a least-squares misfit function.
The FRgWI method is more general as no assumption is 
needed to characterize the passive source: 
The method supports all types of passive sources (distant or 
near from the area of interest, i.e., plane wave or point-source 
illumination), without requiring a priori specification; we 
illustrate in the experiment with local earthquake data 
(see \cref{fig:illustration:teleseismic}).

The benefits of the elastic reciprocity-gap waveform
inversion comes at the price of taking measurements of both the
displacement \emph{and} 
of certain components of the strain tensor.
While measurements of the displacement are common with (string) geophones
(which can be buried, cf.~\cite{Drijkoningen2006}), the possibility to
measure the strain 
can be envisioned with DAS
\citep{Daley2013,AjoFranklin2015,Lindsey2020}. 
Depending on the configuration of the cable, different components of 
the strain can be retrieved, as highlighted by \cite{Sava2018}; 
see, in particular, their Figure~2.
Helical and straight cables provide the necessary features to extract
the strain components \citep{Innanen2017, Sava2018}. Comparisons
between geophones and DAS data have been carried out
by \cite{Mateeva2013} and \cite{Spikes2019}; 
here, we use both data jointly.
Earthquakes (in particular teleseismic events) have been 
recorded using DAS system by \cite{Lindsey2017,Jousset2018} 
and \cite{AjoFranklin2019}, hence including long-period 
signals with low-frequency contents, see \cite{Lindsey2020}.
The comparison between earthquake data recorded by DAS and 
geophones is further studied by \cite{Wang2018} while
\cite{Lellouch2019} use DAS data from earthquake to estimate 
velocity profiles.
While the recording and use of such data still remain in 
its early stages, our work proposes a conceptual study 
that introduces a new method to fully exploit them.

The novelties of our work are 
(1) the introduction of the reciprocity-gap 
    misfit functional for elastic wave propagation which allows, 
    from the DAS acquisition system, the 
(2) multi-resolution reconstruction using 
observed data from 
(2a) events from passive sources (low-frequency), and
(2b) exploration acquisition (high-frequency).  
We first detail the method, which fundamental feature is 
to require only little a-priori knowledge regarding 
the sources. Then we demonstrate the performance of an approach 
based on our functional with a computational experiment. 
Here, the data from regional events are first used to recover the 
coarse-scale background model. 
Next, the exploration data are used to retrieve the finer scales. 
While we need the knowledge of the elastic parameters at and
very near the surface (where the optical fibers are buried), 
existing exploration data can be used to recover 
these, thus further strengthening the interplay between 
regional earthquakes and exploration data.

\section{Methodology}
\label{section:method}

We consider the non-linear seismic imaging problem from measurements 
of waves and rely on an iterative minimization algorithm 
for the recovery of sub-surface elastic properties.
We introduce a misfit functional based on the \emph{reciprocity-gap} 
formula, which uses combinations of measurements and simulations.
It uses measurements of the strain, and enables the use of data 
from arbitrary probing sources. 
Note that while we formulate the method in the frequency domain, it 
can be similarly applied in the time domain.

\subsection{Modeling the Data}

We consider the propagation of time-harmonic waves in an elastic 
medium, given in terms of the vectorial displacement field 
$\displacement$ and of the stress tensor of order two $\tensS$ such
that, at frequency $\omega$ and for an (internal) source $\fsrc$, we have
\begin{equation} \label{eq:waves}
  \divergence \tensS(\bx,\omega) \,+\, \omega^2 \rho(\bx) \, \displacement(\bx,\omega) \,= \, \fsrc(\bx)
  \, , \qquad \text{in the domain $\Omega$.}
\end{equation}
Here, $\rho$ is the material density and $\bx$ represents 
the space coordinates. 
In linear elasticity, the stress tensor is related to the 
strain tensor $\tensE$ by Hooke's law:
\begin{equation} \label{eq:hookes}
  \tensS(\bx,\omega) \, = \, \tensC(\bx,\omega) \, \tensE(\bx,\omega) \, ,
  \qquad \text{with} \quad
  \tensE = \dfrac{1}{2} \Big( \nabla \displacement + (\nabla \displacement)^T \Big), 
\end{equation}
where $^T$ indicates the transpose of a matrix.
The physical properties of the medium are encoded in the 
elasticity tensor of order four $\tensC$ which, in the 
case of isotropy, reduces to the Lam\'e parameters 
$\lambda$ and $\mu$, such that \cref{eq:hookes} writes as,
\begin{equation} \label{eq:hookes-law}
  \tensS = \lambda \, Tr(\tensE) \, I_d  \, +  \,  2 \, \mu \, \tensE \,,
\end{equation}
where $Tr$ denotes the trace, and $I_d$ the identity matrix.
We further recall the P- and S-wave speeds, respectively $c_p$ and $c_s$,
as a function of the Lam\'e parameters:
\begin{equation} \label{eq:wave-speeds}
  c_p = \sqrt{(\lambda + 2\mu)/{\rho}} \, , \qquad\qquad c_s = \sqrt{\mu/\rho}.
\end{equation}

At the boundary of the domain $\Omega$, we consider a 
free surface between the air and the 
ground where, in the absence of a source, $\tensS \cdot \n  = 0$. 
We impose absorbing boundary conditions at the other boundaries
(see \cref{fig:illustration:exploration}), we refer to, 
e.g.,~\cite{Givoli1990} and \cite{Higdon1991}.

The forward problem $\forward$ is defined as
a mapping from the model to the data, that is, it gives 
the solution to the wave equation at the location 
of the receivers for a given physical model  
$\bm$ (for instance, $\bm = (\lambda, \, \mu, \, \rho)$ 
for elastic isotropy).
We write the forward problem in terms of the 
displacement and of the normal stress (traction) and 
denote by $X_{\text{rcv}}$ the receiver line.
Considering $\fsrc_k$ the source used for the simulations, 
we define at frequency $\omega$,
\begin{empheq}[left=]{align} \label{eq:forward}
\begin{split}
& \forward (\fsrc_k, \, \bm, \, \omega)  = 
  \Big\{\,\, \simU(\fsrc_k, \, \bm, \, \omega) \, ; \, \, \, 
             \simS(\fsrc_k, \, \bm, \, \omega) \Big\} \\
&  = \Big\{\,\,
    \displacement(\bx, \, \fsrc_k, \, \bm, \, \omega)  \,; \quad
    \tensS(\bx, \, \fsrc_k, \, \bm,\, \omega) \cdot \n \,; \quad  
     \bx \in X_{\text{rcv}}
  \,\,\Big\} \, .
\end{split} \end{empheq}
The index notations $\simU$ and $\simS$ 
refer to the displacement or the normal
stress data respectively, and $\n$ is 
the \emph{principal} normal vector.
The normal stress further relates to the observables 
in DAS acquisition (the strain) via the Hooke's law 
in \cref{eq:hookes} (see the dedicated section below).

\subsection{Full Reciprocity-gap Waveform Inversion (FRgWI): Misfit Functional}
\label{subsection:FRgWI}

The quantitative reconstruction of the subsurface elastic 
properties is recast as an iterative minimization problem 
following the FWI approach and we design a specific misfit 
functional based upon the \emph{reciprocity gap}. 
We first write the functional in terms of the stress 
tensor, and equivalently use the strain data to relate 
to the DAS acquisition system in the next section.

We denote by $\obsU(\gsrc)$ and $\obsSn(\gsrc)$ the measurements 
associated to a source $\gsrc$, for the displacement and the 
normal stress respectively. 
That is, $\gsrc$ and $\fsrc$ (from \cref{eq:forward}) 
stand for the sources (right-hand 
sides of \cref{eq:waves}) that generate the observations and 
the simulations, respectively. 
It means that $\gsrc$ represents the ``real'' or ``observational'' 
sources of the measured events (e.g., a passive source),  
which can be unknown.
On the other hand, $\fsrc$ in \cref{eq:forward} stands for
the ``artificial'' or ``computational'' sources, used for 
the simulations, and their positions are chosen independently.
It means that both set of sources ($\gsrc$ and $\fsrc$) 
can, but \emph{do not have to} be the same.

Omitting the space dependency for clarity, the reciprocity-gap misfit 
functional in terms of the stress is defined such that, 
\begin{equation}\label{eq:misfit_stress}
  \misfit_\sigma(\bm,\omega) = \dfrac{1}{2} 
                  \sum_{j=1}^{n_{\text{src}}^{\text{sim}}} 
                  \sum_{k=1}^{n_{\text{src}}^{\text{obs}}} 
   \bigg\Vert   
     \int_{X_{\text{rcv}}}  \hspace*{-1em}\big(
          \displacement(\fsrc_j, \bm, \omega) \cdot 
           \obsSn(\gsrc_k,\omega) 
  -  
 \obsU(\gsrc_k,\omega) \cdot \big(\tensS(\fsrc_j, \bm, \omega) \cdot \n \big)
 \big) \mathrm{d} \bx \bigg\Vert^2 \hspace*{-.25em} ,
\end{equation}
where $\Vert \cdot \Vert^2$ is the complex $L^2$ inner product.
Because of the correlation of an observation $\data$ and a simulation ($\displacement$, $\tensS$)
in the misfit, the real sources $\gsrc$ are not explicitly 
needed to construct the functional. 
It allows for the independent selection of the computational 
sources $\fsrc$, and for a different number of observational 
sources, $n_{\text{src}}^{\text{obs}}$, compared to the number 
of computational ones, $n_{\text{src}}^{\text{sim}}$.
Then, the two sums in \cref{eq:misfit_stress} imply that each of 
the real source is tested against all of the computational ones.
For the discretization of the integral along the line of 
receivers, we can use a sum over discrete receiver locations or, 
for different configurations, a weighted sum (e.g., 
quadrature rule), see \cite{Montagner2012}.
The misfit functional is derived from the Green's identity, 
which allows to replace the surface integral (along the line 
of receivers) by a volume one (introducing the divergence of 
the stress). 
Then, using the state \cref{eq:waves,eq:hookes}, the functional relates
to the comparison of the physical properties in the interior;
we refer to the derivation given in the Appendix~A of \cite{Faucher2019FRgWI} for more details.

The essence of this approach is two fold: first, 
it does not compare the observations and simulations 
directly but instead works with their cross-correlations 
(which, in the frequency domain, amounts to a multiplication).
Secondly, each product is made of different fields (i.e., displacement and strain). 
Consequently, the set of sources for the data ($\gsrc_k$)
and for the simulations ($\fsrc_j$) is separated.
The positions of the sources that generate the measurements 
are not needed to define the misfit and the simulations 
can use arbitrary sources (in terms of their positions and functions).

\subsection{From DAS acquisition to Reciprocity-gap Inversion}
\label{subsection:equivalence_strain}

Our misfit functional in \cref{eq:misfit_stress} is defined 
in terms of the displacement and of the normal stress. 
However, the strain can be retrieved from DAS 
acquisition system \citep{Sava2018}.
Therefore, we have to rewrite \cref{eq:misfit_stress} in terms 
of the strain.
It suffices to replace the stress in \cref{eq:misfit_stress}
using Hooke's law \cref{eq:hookes}. 
Nonetheless, it means that 
the medium parameters (contained in $\tensC$) must be known 
at the location of the receivers, and that all the strain
components must be measured.

In the case of elastic isotropy where the Hooke's law \cref{eq:hookes-law} 
prevails, the component of the normal stress in the direction $\inddim=\{x,y,z\}$ is given by, 
\begin{equation} \label{eq:equivalence:normal_iso}
  [ \, \tensS \cdot \n \, ]_\inddim
  \, = \, 2 \, \mu \sum_j \epsilon_{\inddim j} \, \n_j \, + \, 
       \lambda \, Tr(\tensE) \, \n_\inddim \, 
  \, = \, 2 \, \mu \, [ \, \tensE \cdot \n \, ]_\inddim     \, + \, 
       \lambda \, \Big( \sum_j \, \epsilon_{jj} \Big) \, \n_\inddim \, .
\end{equation}
To write the reciprocity-gap formula with the 
normal stress components we require, in terms of the strain (i.e., 
the observables in DAS acquisition),
\begin{enumerate}
\item measurements of the normal strain $\tensE\cdot\n$, 
\item measurements of the trace of the strain tensor 
     (or the diagonal coefficients $\epsilon_{jj}$),
\item the values of $\lambda_0(\bx)$ and $\mu_0(\bx)$
      at the position of the receivers $\bx \in X_{\text{rcv}}$.
\end{enumerate}

Replacing the normal stress in \cref{eq:misfit_stress}
with the strain using \cref{eq:equivalence:normal_iso}, 
we can equivalently use the misfit functional in terms
of the strain, which amounts, under isotropy, to
\begin{empheq}[left=]{align} \label{eq:misfit_strain}
\begin{split}
   \misfit(\bm) = \dfrac{1}{2} ~
                  \sum_{j=1}^{n_{\text{src}}^{\text{sim}}} ~
                  \sum_{k=1}^{n_{\text{src}}^{\text{obs}}} ~
  ~ \bigg\Vert ~
&    \int_{X_{\text{rcv}}}  \bigg(
 \displacement(\fsrc_j, \bm) \cdot 
   \Big(
      2\mu_0 \, \obsE(\gsrc_k) \cdot \n 
     \, + \, \lambda_0 \, Tr\big(\obsE(\gsrc_k) \big) \, \n \Big) \\
     &  - \, \obsU(\gsrc_k) \cdot 
   \Big(
      2\mu_0 \, \tensE(\fsrc_j, \bm) \cdot \n 
   \, + \, \lambda_0  \, Tr\big(\tensE(\fsrc_j, \bm) \big) \, \n
   \Big)
 \bigg) \mathrm{d}\, \bx \bigg\Vert^2,
\end{split} \end{empheq}
where $\lambda_0$ and $\mu_0$ are the known Lam\'e parameters
at the position of the receivers, and $\obsE(\gsrc)$ refers to
the measurements of the strain associated to a source $\gsrc$.

\subsection{Combined Data Inversion}
\label{subsection:combination}

Data obtained for seismic exploration can be 
generated by vibroseis trucks, explosions or air guns.
It usually consists in several hundreds or thousands 
of independent point-sources
at the surface, giving reflection data in general. 
In the case of regional earthquakes or teleseismic events,
the source can be several to hundreds of kilometers below the Earth's
surface and 
such events are not controlled. 
In addition, while the source of an earthquake
is characterized by the moment tensor, 
the vibroseis truck imposes a Neumann boundary 
condition (via the traction), see, 
e.g., \cite{Baeten1989,Aki2002,Carcione2007} and \cite{Shi2019}.
Let us first note that the FRgWI method does not require 
the characterization of the observational (real) sources 
and one can use a dense set of computational 
sources to compensate for a sparse observational 
set, as highlighted by \cite{Faucher2019FRgWI}. 

In exploration seismic, the peak frequency 
of the source lays, at least, in the \num{15}--\num{20} \si{\Hz} range,
resulting in unusable (noisy) low-frequency content.
On the contrary, the data from earthquakes
contain signal of very low frequencies (even below 1 \si{\milli\Hz}). 
Therefore, the iterative minimization is conducted 
following these two steps:
\begin{enumerate}
  \item we minimize $\misfit$ using 
        data from regional events for $\obsU$ and $\obsE$.
        This corresponds to a few sources relatively 
        far from the domain of interest, but where the 
        low-frequency content is usable.
  \item From the low-frequency model built after step 1, 
        we minimize $\misfit$ using 
        the exploration data for $\obsU$ and $\obsE$.
        Here, the acquisition is denser and the frequency 
        content higher, to recover the finer details of the models.
\end{enumerate} 
It is crucial that, numerically speaking, steps 1 
and 2 do not require a different computational 
domain. 
FRgWI works with arbitrary observational sources and, 
here, they are taken outside of the computational area. 
That is, the computational domain only consists of the 
exploration part, 
as illustrated in \cref{fig:illustration}. 
As an alternative, one can perform local updates
after a computation on the global domain, e.g., \cite{Robertsson2000} and \cite{Masson2017}.

\subsection{Gradient Computation}

The computation of the gradient of the misfit 
functional with respect to the parameters 
is based on the adjoint-state method, derived 
in the work of \cite{Lions1971} and \cite{Glowinski1985}, 
and implemented by \cite{Chavent1974}; 
it is reviewed in the context of geophysics 
by \cite{Plessix2006}. 
In the adjoint-state method, the gradient of the misfit 
is computed from the adjoint of the forward problem with specific 
right-hand sides: the \emph{backward} problem. 
These new right-hand sides are further referred to as the `\emph{adjoint sources}'.
The method is also at the heart of the adjoint-tomography
technique in seismology \citep{Tromp2005,Fichtner2006adjointa,Fichtner2006adjointb,Tape2007,Bozdaug2016}.
It can also be used for the computations of second-order derivatives, see, e.g., 
\cite{Wang1992,Fichtner2011hessian} and \cite{Metivier2013}.

For the sake of conciseness, we shall only detail the 
adjoint-sources, which are specific to the misfit 
functional, and refer the readers to, e.g., 
\cite{Pratt1998,Plessix2006,Chavent2010,Alessandrini2019} 
and \cite{Faucher2019IP} for details of the method in seismic applications.
The adjoint-sources are given by the 
derivatives of the misfit functional with respect to each
component of the wavefield (here, $\displacement$ and 
$\tensS$). 
For each of the \emph{computational} sources 
$\fsrc_j$ in \cref{eq:misfit_stress} corresponds a backward 
problem where each of the unknowns has a right-hand side $W_\bullet$,
where $\bullet$ refers to the component.
Using $\inddim=\{x,y,z\}$ to denote the direction 
(e.g., $W_{u_x}$ is associated with $u_x$), 
the right-hand sides are given by, for $\bx \in X_{\text{rcv}}$,
\begin{empheq}[left={\empheqlbrace}]{align} \begin{split}
  W_{u_{\inddim}}(\fsrc_j, \bx)  \, & := \, \phantom{+} \, 
  \sum_{k=1}^{n_{\text{src}}^{\text{obs}}} \, \overline{\eta}(\fsrc_j,\gsrc_k) \,
                   \big( \obsS(\gsrc_k, \bx) \big)_{\inddim} \, ,
                   \\
  W_{\tensS_{\inddim\inddim}}(\fsrc_j, \bx)  & :=  -  
  \sum_{k=1}^{n_{\text{src}}^{\text{obs}}}  \overline{\eta}(\fsrc_j,\gsrc_k)  
                   \big( \obsU(\gsrc_k, \bx) \big)_{\inddim}  \n_\inddim \, 
                   \, , \\
  W_{\tensS_{\inddim_1 \inddim_2}}
    (\fsrc_j, \bx)  & :=  -  
    \sum_{k=1}^{n_{\text{src}}^{\text{obs}}}  \overline{\eta}(\fsrc_j,\gsrc_k)  
                     \sum_{\inddim_1 \neq \inddim_2}  
                     \big( \obsU(\gsrc_k, \bx) \big)_{\inddim_1}  \n_{\inddim_2}
                     \, , \quad \text{for $\inddim_1 \neq \inddim_2$,}
\end{split} \end{empheq}
with
\begin{equation}
  {\eta}(\fsrc_j,\gsrc_k) \, = \, \int_{X_{\text{rcv}}} \bigg( 
      \displacement(\fsrc_j, \bx, \bm) \cdot \obsS(\gsrc_k, \bx)
  \, -  \, \obsU(\gsrc_k, \bx) \cdot \big(\tensS(\bx, \fsrc_j, \bm)  \cdot \n \big) \,\bigg)\, \mathrm{d}\bx \, ,
\end{equation}
using $\overline{\cdot}$ to denote the complex conjugation.
We see that each of the adjoint-sources take the contribution 
from \emph{all} the measurement sources ($\gsrc_k$) that is, 
from all the observed data.
Eventually, the gradient is obtained combining the forward and 
backward solutions \citep{Pratt1998}.

  The reciprocity-gap misfit functional can also be written 
  from data at the surface, for instance in the exploration 
  settings where the source is the traction, $\tensS \cdot \n$, 
  imposed by the vibroseis \citep{Baeten1989}.
  In this case, with displacement data acquired by surface 
  geophones, 
  it gives the Neumann-to-Dirichlet map \citep{Shi2019}, 
  which graph forms the necessary Cauchy data for reciprocity gap
  (i.e., displacement and normal stress).
  Then, the misfit functional in~\cref{eq:misfit_stress} is written by
  replacing the values of the normal stress by the 
  imposed source traction, such that,
 \begin{equation} \label{eq:misfit_stess:surface}
   \misfit_\text{surface}(\bm) = \dfrac{1}{2} \,
                   \sum_{j=1}^{n_{\text{src}}^{\text{sim}}}
                   \sum_{k=1}^{n_{\text{src}}^{\text{obs}}} \,
   ~ \bigg\Vert \,
      \int_{X_{\text{rcv}} \subset \Gamma}  \hspace*{-1mm} \big(
          \displacement(\fsrc_j, \bx, \bm) \cdot 
            \gsrc_k(\bx)
 \, - \, \obsU(\gsrc_k, \bx) \cdot \fsrc_j(\bx)
  \big) \mathrm{d}\, \bx \bigg\Vert^2 .
 \end{equation}
  To obtain the gradient, the backward problem correspond to 
  a boundary value problem where the adjoint-sources,
  following the steps prescribed by \cite{Shi2019}, are
 \begin{empheq}[left=]{align} \begin{split}
    \tensS \, \cdot \, \n \, = \, &
  \sum_{k=1}^{n_{\text{src}}^{\text{obs}}} \, \gsrc_k(\bx) \, \overline{
      \int_{X_{\text{rcv}} \subset \Gamma}  \hspace*{0mm}\, \Big(
      \displacement(\fsrc_j, \bx, \bm) \cdot \gsrc_k(\bx)
   \, -  \, \obsU(\gsrc_k, \bx) \cdot \fsrc_j(\bx) \, \Big) \, \mathrm{d} \bx} \, , 
  \end{split} \end{empheq} 
  Nonetheless, as the source traction is imposed only 
  at the position of the vibroseis base-plate (equal to zero elsewhere),  
  it is unclear if it performs well.
  In the following experiment, we consider, per convenience, 
  buried devices for the exploration acquisition setting. 

\subsection{Numerical implementation}

To simulate the displacement and the stress tensor (or 
the strain), we implement the hybridizable discontinuous 
Galerkin (HDG) method \citep{Arnold2002,Cockburn2009} for 
the discretization of \cref{eq:waves}. 
The motivation is that HDG solves the first-order problem 
(i.e., the system made of~\cref{eq:waves,eq:hookes}), hence 
gives access to both the displacement and stress tensor,
while generating a linear system relatively small compared
to other discretization approaches \citep{Bonnasse2017,Faucher2020adjoint}.

The HDG method works in two steps. In the first step, the \emph{global}
linear system, whose matrix only accounts for the degrees 
of freedoms (dof) of the numerical trace for one variable (the displacement).
In the second step, local (for each cell of the mesh), 
small systems are  solved to have, from the numerical trace
obtained at the global stage, the volume solutions for the displacement
\emph{and} for the stress tensor \citep{Bonnasse2017}.
In other discretization methods such as finite elements or internal 
penalty discontinuous Galerkin, upon discretization of the first-order problem
\cref{eq:waves,eq:hookes}, the resulting linear system is of size 
the number of dof for \emph{all} unknowns (the stress tensor and the displacement).
On the other hand, with the HDG method, the global linear system is 
smaller as it only takes the dof of the trace of one of the unknowns (the displacement).
It makes it an appropriate choice for frequency-domain applications, 
where the bottleneck usually is the computational memory required to 
factorize the matrix and solve the linear system.
We further refer to \cite{Kirby2012,Bonnasse2017} and 
\cite{Faucher2020adjoint} for more details on the HDG 
discretization and its performance.
As it gives access to both solutions (stress and displacement)
that are required for the reciprocity-gap misfit functional, 
the HDG method is also used in the acoustic settings of FRgWI 
by \cite{Faucher2019FRgWI}.

\section{Computational Experiment}
\label{section:experiments}

We illustrate the performance of FRgWI with a two-dimensional 
isotropic elastic experiment where we consecutively use exploration 
and regional earthquake  data. 
The data are generated using a domain of size 
$28\times 5$ \si{\kilo\meter\squared} and we consider a domain of 
size $22\times 4.5$ \si{\kilo\meter\squared} for the reconstruction, 
as illustrated in \cref{fig:illustration}.
The elastic properties are pictured in 
\cref{fig:frgwi:true_models_rho,fig:frgwi:true_models_cp,fig:frgwi:true_models_cs}, 
with the density, P-wave speed and S-wave speed
respectively, and where we indicate by white dashes the 
restriction to the inversion domain
(from $4.5$ to $22$ \si{\km} in $x$ and from $0$ 
to $4.5$ \si{\km} in depth). 
The model is composed of a body of high contrast in its 
center, with layered structures on the sides and below.
In addition, there is a strong contrast in 
speeds between the layers.

\pgfmathsetmacro{\xmingb}  {0.00}
\pgfmathsetmacro{\xmaxgb} {28.00}
\pgfmathsetmacro{\zmingb}  {0.00}
\pgfmathsetmacro{\zmaxgb}  {5.00}
\pgfmathsetmacro{\xminloc} {4.50}
\pgfmathsetmacro{\xmaxloc}{22.00}
\pgfmathsetmacro{\zminloc} {0.00}
\pgfmathsetmacro{\zmaxloc} {4.00}

\setlength{\modelwidth}  {7.00cm} 
\setlength{\modelheight} {4.25cm}
\setlength{\modelRwidth} {5.00cm} 
\setlength{\modelRheight}{3.75cm}
\setlength{\rbox}         {7mm}
\setlength{\myhspace}    {-2mm}
\pgfmathsetmacro{\capx} {-0.25}
\pgfmathsetmacro{\capy}  {3.15}
\pgfmathsetmacro{\capRy} {2.75}

\setlength{\modelwidth}  {7.70cm} 
\setlength{\modelheight} {4.25cm}
\setlength{\modelRwidth} {5.30cm} 
\setlength{\modelRheight}{3.75cm}
\setlength{\rbox}         {5.0mm}
\setlength{\myhspace}    {-2mm}
\pgfmathsetmacro{\capx} {-0.25}
\pgfmathsetmacro{\capy}  {3.25}
\pgfmathsetmacro{\capRy} {2.75}
\begin{figure}[ht!] \centering
  \renewcommand{\modelfile}{rho_true}  \renewcommand{\mysubcaption}{\textbf{a)}}
  \subfloat{\label{fig:frgwi:true_models_rho} \begin{tikzpicture}

\pgfmathsetmacro{\cmin}   {0.50}
\pgfmathsetmacro{\cmax}   {2.75}

\begin{axis}[%
  width =\modelwidth,
  height=\modelheight,
  axis on top, separate axis lines,
  xmin=\xmingb, xmax=\xmaxgb, xlabel={$x$   (\si{\km})},
  ymin=\zmingb, ymax=\zmaxgb, ylabel={depth (\si{\km})}, y dir=reverse,
  xlabel style={yshift=2mm, xshift=0mm},
  colormap/jet,colorbar,
  ytick={0,2,4},
  colorbar style={title={$\rho$ (\si{\kg\per\meter\cubed})},
  title style={yshift=-3mm, xshift=-10mm},
  xshift=-.2cm, width=.2cm,
  },point meta min=\cmin,point meta max=\cmax
]
\addplot [forget plot] graphics [xmin=\xmingb,xmax=\xmaxgb,ymin=\zmingb,ymax=\zmaxgb] 
                                {{\modelfile}.png};

\draw[line width=2pt, dashed, white] (\xminloc,\zminloc) -- (\xmaxloc,\zminloc) -- 
                                     (\xmaxloc,\zmaxloc) -- (\xminloc,\zmaxloc) -- cycle;

\end{axis}
\node[text width=2cm,anchor=north west,black] at (\capx,\capy) {\mysubcaption};
\end{tikzpicture}%
}
  \hspace*{\myhspace} 
  \renewcommand{\modelfile}{rho_start}  \renewcommand{\mysubcaption}{\textbf{b)}}
  {\raisebox{\rbox}{
  \subfloat{\label{fig:frgwi:start_models_rho} \begin{tikzpicture}

\pgfmathsetmacro{\cmin}   {0.50}
\pgfmathsetmacro{\cmax}   {2.75}

\begin{axis}[%
  width =\modelRwidth,
  height=\modelRheight,
  axis on top, separate axis lines,
  xmin=\xminloc, xmax=\xmaxloc, xlabel={$x$   (\si{\km})},
  ymin=\zminloc, ymax=\zmaxloc, 
  y dir=reverse,
  ytick={0,2,4},
  xlabel style={yshift=2mm},
  colormap/jet,colorbar,
  colorbar style={
    title={$\rho$ (\si{\kg\per\meter\cubed})},
    title style={yshift=-3mm, xshift=-10mm},
    xshift=-.2cm, width=.2cm,
    },point meta min=\cmin,point meta max=\cmax
]
\addplot [forget plot] graphics [xmin=\xmingb,xmax=\xmaxgb,ymin=\zmingb,ymax=\zmaxgb] 
                                {{\modelfile}.png};

\end{axis}
\node[text width=2cm,anchor=north west,black] at (\capx,\capRy) {\mysubcaption};
\end{tikzpicture}%
}}}
  \\ 
  \renewcommand{\modelfile}{cp_true}  \renewcommand{\mysubcaption}{\textbf{c)}}
  \subfloat{\label{fig:frgwi:true_models_cp} \begin{tikzpicture}

\pgfmathsetmacro{\cmin}   {2.50}
\pgfmathsetmacro{\cmax}   {5.50}

\begin{axis}[%
  width =\modelwidth,
  height=\modelheight,
  axis on top, separate axis lines,
  xmin=\xmingb, xmax=\xmaxgb, xlabel={$x$   (\si{\km})},
  ymin=\zmingb, ymax=\zmaxgb, ylabel={depth (\si{\km})}, y dir=reverse,
  xlabel style={yshift=2mm, xshift=0mm},
  colormap/jet,colorbar,
  ytick={0,2,4},
  colorbar style={title={$c_p$ (\si{\km\per\second})},
  title style={yshift=-3mm, xshift=-10mm},
  xshift=-.2cm, width=.2cm,
  },point meta min=\cmin,point meta max=\cmax
]
\addplot [forget plot] graphics [xmin=\xmingb,xmax=\xmaxgb,ymin=\zmingb,ymax=\zmaxgb] 
                                {{\modelfile}.png};

\draw[line width=2pt, dashed, white] (\xminloc,\zminloc) -- (\xmaxloc,\zminloc) -- 
                                     (\xmaxloc,\zmaxloc) -- (\xminloc,\zmaxloc) -- cycle;
\end{axis}
\node[text width=2cm,anchor=north west,black] at (\capx,\capy) {\mysubcaption};

\end{tikzpicture}%
}
  \hspace*{\myhspace} 
  \renewcommand{\modelfile}{cp_start}  \renewcommand{\mysubcaption}{\textbf{d)}}
  {\raisebox{\rbox}{
  \subfloat{\label{fig:frgwi:start_models_cp} \begin{tikzpicture}

\pgfmathsetmacro{\cmin}   {2.50}
\pgfmathsetmacro{\cmax}   {5.50}

\begin{axis}[%
  width =\modelRwidth,
  height=\modelRheight,
  axis on top, separate axis lines,
  xmin=\xminloc, xmax=\xmaxloc, xlabel={$x$   (\si{\km})},
  ymin=\zminloc, ymax=\zmaxloc, 
  xlabel style={yshift=2mm},
  y dir=reverse,
  ytick={0,2,4},
  colormap/jet,colorbar,
  colorbar style={title={$c_p$ (\si{\km\per\second})},
  title style={yshift=-3mm, xshift=-10mm},
  xshift=-.2cm, width=.2cm,
  },point meta min=\cmin,point meta max=\cmax
]
\addplot [forget plot] graphics [xmin=\xmingb,xmax=\xmaxgb,ymin=\zmingb,ymax=\zmaxgb] 
                                {{\modelfile}.png};

\end{axis}
\node[text width=2cm,anchor=north west,black] at (\capx,\capRy) {\mysubcaption};
\end{tikzpicture}%
}}}
  \\ 
  \renewcommand{\modelfile}{cs_true}  \renewcommand{\mysubcaption}{\textbf{e)}}
  \subfloat{\label{fig:frgwi:true_models_cs} \begin{tikzpicture}

\pgfmathsetmacro{\cmin}   {1.70}
\pgfmathsetmacro{\cmax}   {3.20}

\begin{axis}[%
  width =\modelwidth,
  height=\modelheight,
  axis on top, separate axis lines,
  xmin=\xmingb, xmax=\xmaxgb, xlabel={$x$   (\si{\km})},
  ymin=\zmingb, ymax=\zmaxgb, ylabel={depth (\si{\km})}, y dir=reverse,
  xlabel style={yshift=2mm, xshift=0mm},
  colormap/jet,colorbar,
  ytick={0,2,4},
  colorbar style={title={$c_s$ (\si{\km\per\second})},
  title style={yshift=-3mm, xshift=-10mm},
  xshift=-.2cm, width=.2cm, ytick={2,3},
  },point meta min=\cmin,point meta max=\cmax
]
\addplot [forget plot] graphics [xmin=\xmingb,xmax=\xmaxgb,ymin=\zmingb,ymax=\zmaxgb] 
                                {{\modelfile}.png};

\draw[line width=2pt, dashed, white] (\xminloc,\zminloc) -- (\xmaxloc,\zminloc) -- 
                                     (\xmaxloc,\zmaxloc) -- (\xminloc,\zmaxloc) -- cycle;
\end{axis}
\node[text width=2cm,anchor=north west,black] at (\capx,\capy) {\mysubcaption};
\end{tikzpicture}%
} 
  \hspace*{\myhspace} 
  \renewcommand{\modelfile}{cs_start}  \renewcommand{\mysubcaption}{\textbf{f)}}
  {\raisebox{\rbox}{
  \subfloat{\label{fig:frgwi:start_models_cs} \begin{tikzpicture}

\pgfmathsetmacro{\cmin}   {1.70}
\pgfmathsetmacro{\cmax}   {3.20}

\begin{axis}[%
  width =\modelRwidth,
  height=\modelRheight,
  axis on top, separate axis lines,
  xmin=\xminloc, xmax=\xmaxloc, xlabel={$x$   (\si{\km})},
  ymin=\zminloc, ymax=\zmaxloc, 
  xlabel style={yshift=2mm},
  y dir=reverse,
  ytick={0,2,4},
  colormap/jet,colorbar,
  colorbar style={title={$c_s$ (\si{\km\per\second})},
  title style={yshift=-3mm, xshift=-10mm},
  xshift=-.2cm, width=.2cm,  ytick={2,3}
  },point meta min=\cmin,point meta max=\cmax,
]
\addplot [forget plot] graphics [xmin=\xmingb,xmax=\xmaxgb,ymin=\zmingb,ymax=\zmaxgb] 
                                {{\modelfile}.png};

\end{axis}
\node[text width=2cm,anchor=north west,black] at (\capx,\capRy) {\mysubcaption};
\end{tikzpicture}%
}}}
  \caption{Target 
           \protect\subref{fig:frgwi:true_models_rho} density
           \protect\subref{fig:frgwi:true_models_cp} P-wave speed
           and \protect\subref{fig:frgwi:true_models_cs} S-wave speed
           models of size $28\times 5$ \si{\km\squared}, from which 
           the observed measurements are generated.
           The white dashes indicate the imaging domain
           of size $22\times 4.5$ \si{\km\squared}, corresponding to 
           the size of the initial 
           \protect\subref{fig:frgwi:start_models_rho} density
           \protect\subref{fig:frgwi:start_models_cp} P-wave speed
           and \protect\subref{fig:frgwi:start_models_cs} S-wave speed.}
  \label{fig:frgwi:true}
\end{figure}

\subsection{Multi-resolution reconstruction}

For the reconstruction of the elastic physical properties, 
we follow the workflow described above, minimizing the 
misfit function $\misfit$ of \cref{eq:misfit_stress} and using both exploration 
and regional earthquake data. 
\begin{enumerate}
  \item In the first stage, we use data from 
        regional events.
        There are ten passive sources (${n_{\text{src}}^{\text{obs}}}=10$), 
        whose positions are outside of the computational domain
        (\cref{fig:illustration:teleseismic}).
        This is a relatively small data set 
        but it contains usable low-frequencies, 
        and we use contents from 0.2 \si{\Hz} to 2 \si{\Hz}.
  \item In the second stage, we use data from an 
        exploration acquisition setup,
        with ${n_{\text{src}}^{\text{obs}}}=\num{89}$ 
        sources located at the surface.
        This data set is dense but does not contain 
        low-frequencies below $2$ \si{\Hz}.
\end{enumerate} 
Despite using synthetic data, we incorporate 
white Gaussian noise in the measurements
and use different meshes and order of polynomials 
between the generation of synthetic data and the inversion 
scheme.
We assume the measurements are made by 359 receivers, 
located below the surface,
every 50 \si{\meter}. 
The initial models for the reconstruction are shown 
in \cref{fig:frgwi:start_models_cp,fig:frgwi:start_models_cs,fig:frgwi:start_models_rho}.
In our experiment, we further assume the knowledge of the 
near-surface area where the parameters are taken constant
(see \cref{fig:illustration}). 
This allows us to reduce the effect of the free-surface 
condition that generates reflections, increasing 
the non-linearity of the inversion procedure \citep{Brossier2009}.
As an alternative, the use of a regularization term or 
a smoothing filter applied onto the gradient can 
also be used to
balance the contributions of the free-surface 
\citep{Guitton2012,Trinh2017}.

With FRgWI, the choice of the sources for the computational 
acquisition is arbitrary (compared to more traditional misfit 
criterion which must respect the observational sources)
and we take $n_{\text{src}}^{\text{sim}}=\num{89}$ 
computational sources, similarly to the exploration setup, 
for simplicity only. 
We refer to \cite{Faucher2019FRgWI} for more details on the 
flexibility in the choice and influence of these numerical 
sources, where the efficiency with respect to shot summation
is investigated.
In \cref{fig:frgwi:gradient_tele,fig:frgwi:gradient_explo},
we compare the gradient of the misfit functional with respect
to the parameter $1/\mu$ for the regional earthquake  and 
the exploration data set, using the starting models 
of \cref{fig:frgwi:true}, where we force to zero the upper (known) area. 
In both cases, the acquisition for the simulation (which 
is arbitrary with FRgWI) is using sources near the surface 
only, as mentioned above. 
For the regional earthquake data, we use a frequency of \num{0.2} \si{\Hz}: 
we see in \cref{fig:frgwi:gradient_tele} that the gradient 
shows long wavelength variations. 
We note higher amplitudes on the sides and 
bottom regions, that is, near where the passive sources are located,
even if the computational ones are positioned near the surface. 
With the exploration data set (\cref{fig:frgwi:gradient_explo})
we use a frequency of $2$ \si{\Hz}: we observe that the structures 
are smaller (higher wavenumber), with higher amplitudes in the 
upper part of the domain, where the exploration sources are
located.

\begin{figure}[ht!] \centering
  \pgfmathsetmacro{\cmin}  {-0.1}
  \pgfmathsetmacro{\cmax}   {0.1}
  \renewcommand{\modelfile}{search-dir-mu_0.2hz_scale-0.1to0.1} \renewcommand{\mysubcaption}{\textbf{a)}}
  \subfloat{\label{fig:frgwi:gradient_tele} \begin{tikzpicture}

\begin{axis}[%
  width =\modelRwidth,
  height=\modelRheight,
  axis on top, separate axis lines,
  xmin=\xminloc, xmax=\xmaxloc, xlabel={$x$   (\si{\km})},
  ymin=\zminloc, ymax=\zmaxloc, ylabel={depth (\si{\km})}, 
  xlabel style={yshift=2mm},
  y dir=reverse,
  ytick={0,2,4},
  colormap/jet,colorbar,
  colorbar style={title={$\nabla_{1/\mu} \misfit$ ~~(\si{\newton\cubed\per\meter\squared})},
  title style={yshift=-2mm, xshift=-7mm},
  xshift=-.2cm, width=.2cm, ytick={\cmin,0,\cmax},
  },point meta min=\cmin,point meta max=\cmax
]
\addplot [forget plot] graphics [xmin=\xmingb,xmax=\xmaxgb,ymin=\zmingb,ymax=\zmaxgb] 
                                {{\modelfile}.png};

\end{axis}
\node[text width=2cm,anchor=north west,black] at (\capx,\capRy) {\mysubcaption};
\end{tikzpicture}%
} 
  \pgfmathsetmacro{\cmin}  {-0.4}
  \pgfmathsetmacro{\cmax}   {0.4}
  \renewcommand{\modelfile}{search-dir-mu_2hz_scale-0.4to0.4} \renewcommand{\mysubcaption}{\textbf{b)}}
  \subfloat{\label{fig:frgwi:gradient_explo} \begin{tikzpicture}

\begin{axis}[%
  width =\modelRwidth,
  height=\modelRheight,
  axis on top, separate axis lines,
  xmin=\xminloc, xmax=\xmaxloc, xlabel={$x$   (\si{\km})},
  ymin=\zminloc, ymax=\zmaxloc, ylabel={depth (\si{\km})}, 
  xlabel style={yshift=2mm},
  y dir=reverse,
  ytick={0,2,4},
  colormap/jet,colorbar,
  colorbar style={title={$\nabla_{1/\mu} \misfit$ ~~(\si{\newton\cubed\per\meter\squared})},
  title style={yshift=-2mm, xshift=-7mm},
  xshift=-.2cm, width=.2cm, ytick={\cmin,0,\cmax},
  },point meta min=\cmin,point meta max=\cmax
]
\addplot [forget plot] graphics [xmin=\xmingb,xmax=\xmaxgb,ymin=\zmingb,ymax=\zmaxgb] 
                                {{\modelfile}.png};

\end{axis}
\node[text width=2cm,anchor=north west,black] at (\capx,\capRy) {\mysubcaption};
\end{tikzpicture}%
}
  \caption{
           Comparison of the FRgWI gradient for the parameter $1/\mu$
           \protect\subref{fig:frgwi:gradient_tele} using the regional earthquake 
           data ($n_{\text{src}}^{\text{obs}}=\num{10}$  sources) 
           at \num{0.2} \si{\Hz} 
           and \protect\subref{fig:frgwi:gradient_explo} using the exploration
           data ($n_{\text{src}}^{\text{obs}}=\num{89}$ sources) 
           at \num{2} \si{\Hz}; both use the same acquisition
           for the simulations ($n_{\text{src}}^{\text{sim}}=\num{89}$ sources)
           and we force the values of the gradient to be zero in the layer 
           where the parameters are known, according to \cref{fig:illustration}.
           }
  \label{fig:frgwi:gradient}
\end{figure}

\begin{figure}[ht!] \centering
  \renewcommand{\modelfile}{cp_10hz_exploration_after-teleseismic-recycle}  
  \renewcommand{\mysubcaption}{\textbf{a)}}
  \subfloat{\label{fig:frgwi:explo_cp} \begin{tikzpicture}

\pgfmathsetmacro{\cmin}   {2.50}
\pgfmathsetmacro{\cmax}   {5.50}

\begin{axis}[%
  width =\modelRwidth,
  height=\modelRheight,
  axis on top, separate axis lines,
  xmin=\xminloc, xmax=\xmaxloc, xlabel={$x$   (\si{\km})},
  ymin=\zminloc, ymax=\zmaxloc, ylabel={depth (\si{\km})}, y dir=reverse,
  xlabel style={yshift=2mm},
  ylabel style={yshift=-2mm},
  y dir=reverse,ytick={0,2,4},
  colormap/jet,colorbar,
  colorbar style={title={$c_p$ (\si{\km\per\second})},
  title style={yshift=-2mm, xshift=-10mm},
  xshift=-.2cm, width=.2cm,
  },point meta min=\cmin,point meta max=\cmax
]
\addplot [forget plot] graphics [xmin=\xminloc,xmax=\xmaxloc,ymin=\zminloc,ymax=\zmaxloc] 
                                {{\modelfile}.png};

\end{axis}
\node[text width=2cm,anchor=north west,black] at (\capx,\capRy) {\mysubcaption};
\end{tikzpicture}%
}
  \renewcommand{\modelfile}{cs_10hz_exploration_after-teleseismic-recycle}  
  \renewcommand{\mysubcaption}{\textbf{b)}}
  \subfloat{\label{fig:frgwi:explo_cs} \begin{tikzpicture}

\pgfmathsetmacro{\cmin}   {1.70}
\pgfmathsetmacro{\cmax}   {3.20}

\begin{axis}[%
  width =\modelRwidth,
  height=\modelRheight,
  axis on top, separate axis lines,
  xmin=\xminloc, xmax=\xmaxloc, xlabel={$x$   (\si{\km})},
  ymin=\zminloc, ymax=\zmaxloc, 
  y dir=reverse, 
  xlabel style={yshift=2mm},
  y dir=reverse,ytick={0,2,4},
  colormap/jet,colorbar,
  colorbar style={title={$c_s$ (\si{\km\per\second})},
  title style={yshift=-2mm, xshift=-10mm},
  xshift=-.2cm, width=.2cm,ytick={2,3},
  },point meta min=\cmin,point meta max=\cmax
]
\addplot [forget plot] graphics [xmin=\xminloc,xmax=\xmaxloc,ymin=\zminloc,ymax=\zmaxloc] 
                                {{\modelfile}.png};

\end{axis}
\node[text width=2cm,anchor=north west,black] at (\capx,\capRy) {\mysubcaption};
\end{tikzpicture}%
} 
  \\[-0.4\baselineskip] 
  \renewcommand{\modelfile}{cp_2hz_teleseismic}  
  \renewcommand{\mysubcaption}{\textbf{c)}}
  \subfloat{\label{fig:frgwi:tele_2hz_cp} \begin{tikzpicture}

\pgfmathsetmacro{\cmin}   {2.50}
\pgfmathsetmacro{\cmax}   {5.50}

\begin{axis}[%
  width =\modelRwidth,
  height=\modelRheight,
  axis on top, separate axis lines,
  xmin=\xminloc, xmax=\xmaxloc, xlabel={$x$   (\si{\km})},
  ymin=\zminloc, ymax=\zmaxloc, ylabel={depth (\si{\km})}, y dir=reverse,
  xlabel style={yshift=2mm},
  ylabel style={yshift=-2mm},
  y dir=reverse,ytick={0,2,4},
  colormap/jet,colorbar,
  colorbar style={title={$c_p$ (\si{\km\per\second})},
  title style={yshift=-2mm, xshift=-10mm},
  xshift=-.2cm, width=.2cm,
  },point meta min=\cmin,point meta max=\cmax
]
\addplot [forget plot] graphics [xmin=\xminloc,xmax=\xmaxloc,ymin=\zminloc,ymax=\zmaxloc] 
                                {{\modelfile}.png};

\end{axis}
\node[text width=2cm,anchor=north west,black] at (\capx,\capRy) {\mysubcaption};
\end{tikzpicture}%
}
  \renewcommand{\modelfile}{cs_2hz_teleseismic} 
  \renewcommand{\mysubcaption}{\textbf{d)}}
  \subfloat{\label{fig:frgwi:tele_2hz_cs} \begin{tikzpicture}

\pgfmathsetmacro{\cmin}   {1.70}
\pgfmathsetmacro{\cmax}   {3.20}

\begin{axis}[%
  width =\modelRwidth,
  height=\modelRheight,
  axis on top, separate axis lines,
  xmin=\xminloc, xmax=\xmaxloc, xlabel={$x$   (\si{\km})},
  ymin=\zminloc, ymax=\zmaxloc, 
  y dir=reverse, 
  xlabel style={yshift=2mm},
  y dir=reverse,ytick={0,2,4},
  colormap/jet,colorbar,
  colorbar style={title={$c_s$ (\si{\km\per\second})},
  title style={yshift=-2mm, xshift=-10mm},
  xshift=-.2cm, width=.2cm,ytick={2,3},
  },point meta min=\cmin,point meta max=\cmax
]
\addplot [forget plot] graphics [xmin=\xminloc,xmax=\xmaxloc,ymin=\zminloc,ymax=\zmaxloc] 
                                {{\modelfile}.png};

\end{axis}
\node[text width=2cm,anchor=north west,black] at (\capx,\capRy) {\mysubcaption};
\end{tikzpicture}%
}
  \\
  \subfloat{\label{fig:frgwi:explo_cp_1d} 
            \includegraphics[scale=1]{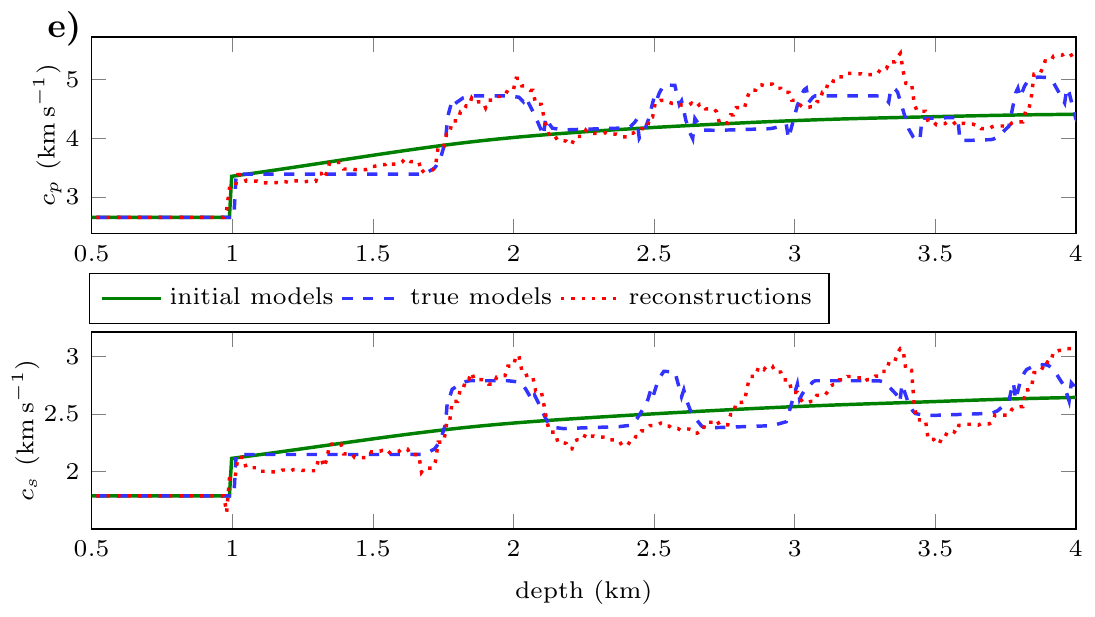}}
  \caption{The reconstructed \protect\subref{fig:frgwi:explo_cp}
           P-wave speed and \protect\subref{fig:frgwi:explo_cs} 
           S-wave speed models use the exploration data from 
           $2$ \si{\Hz} to $10$ \si{\Hz}, starting from the 
           \protect\subref{fig:frgwi:tele_2hz_cp} P-wave speed
           and  \protect\subref{fig:frgwi:tele_2hz_cs}
           S-wave speed models built with the low-frequency 
           data from regional events.
           \protect\subref{fig:frgwi:explo_cp_1d} Vertical sections
           of the target, starting and reconstructed models at 
           $x=16.5$ \si{\km}.}
  \label{fig:frgwi:all}
\end{figure}

We follow a frequency continuation approach
\citep{Bunks1995,Faucher2020Geo}, and perform $30$ 
iterations per frequency.
The first stage (using the data from passive sources)
uses frequencies \num{0.2} \si{\Hz}, \num{0.4} \si{\Hz}, 
\num{0.6} \si{\Hz}, \num{0.8} \si{\Hz}, \num{1} \si{\Hz} 
and \num{2} \si{\Hz}. The second stage with the exploration 
data set uses frequencies from $2$ \si{\Hz} 
to $10$ \si{\Hz}, every $1$ \si{\Hz}.
While the density is not updated, 
this lack of information should not prevent
us from recovering the other parameters and we further 
select the parametrization $1/\lambda$ and $1/\mu$ for the 
inversion \citep{Faucher2017}.
The results after the iterations with the low-frequency 
regional earthquake  data are pictured in 
\cref{fig:frgwi:tele_2hz_cp,fig:frgwi:tele_2hz_cs}, where
we visualize the wave speeds (\cref{eq:wave-speeds}).
Then, these reconstructed models serve as 
initial ones to carry on the reconstruction 
with the exploration data.
The results are shown in \cref{fig:frgwi:explo_cp,fig:frgwi:explo_cs}.

The low-frequency reconstruction provides the 
background profile of the parameters, and is 
able to discover the contrasting body in the 
center. As expected, the parameters are smooth
at this stage, with only the long wavelengths
investigated. 
Despite the use of data from few passive sources located
outside the inverted area, the inversion procedure
gives access to the 
low-frequency profiles of the medium parameters.
In the second stage, the reconstruction of the 
higher wavenumbers 
is obtained from the exploration data. 
Eventually, we see in \cref{fig:frgwi:all} 
that the upper part of the contrast is correctly recovered. 
Some of the underneath layers appear, but the 
deep and side parts of the models are mostly missing 
due to the limited illumination, as highlighted in 
the vertical section in $x=16.5$ \si{\km} in \cref{fig:frgwi:explo_cp_1d}.

\subsection{Comparison of misfit functions in the second stage}

In the first stage with the data from the regional events, 
the FRgWI method is essential as it does not 
require the positions of the passive sources.
However, in the second stage with the exploration data, 
it is common to assume the knowledge of the sources 
(e.g., the position of the vibroseis truck) and thus 
it is possible to rely on a more traditional misfit 
criterion, by evaluating the difference between 
the observed and simulated data as follows:
\begin{equation}\label{eq:misfit_raw}
  \misfit_0(\bm)\, =\, 
          \sum_{k=1}^{n_{\text{src}}^{\text{obs}}} \,
          \dfrac{1}{2} \Big\Vert ~  
          \simU(\gsrc_k, \bm) \,-\, \obsU(\gsrc_k)
          \Big\Vert^2 
        ~+~    
          \dfrac{1}{2} \Big\Vert ~  
          \simE(\gsrc_k, \bm) \,-\, \obsE(\gsrc_k)
          \Big\Vert^2 ,
\end{equation}
where the forward problem is defined in \cref{eq:forward}.
Here, contrary to our misfit function \cref{eq:misfit_strain},
the observational sources $\gsrc$ must be used to generate the 
simulations.

Now, starting from the models recovered after the 
first stage using the low frequencies in 
\cref{fig:frgwi:tele_2hz_cp,fig:frgwi:tele_2hz_cs},
we perform the minimization of $\misfit_0$ 
(\cref{eq:misfit_raw}) using the exploration data set. 
We compare the evolution of the cost functions 
for the first frequency of the exploration data set, 
$2$ \si{\Hz}, in \cref{fig:misfit-comparison:misfit}.
While $\misfit$ allows the flexibility of the computational 
sources ($\fsrc$ in \cref{eq:misfit_strain}), we use the same 
set as the observational ones (i.e., $\gsrc$) to have a consistent 
comparison between the two methods.
In \cref{fig:misfit-comparison:models}, we provide the 
evolution of the relative model error, defined as,
\begin{equation} \label{eq:model-error}
  \mathfrak{e}_k(c, \, \misfit) \, = \, \bigg\Vert \dfrac{c_k(\misfit)  - c^\dagger}{c^\dagger} \bigg\Vert \, ,
\end{equation}
where $c_k(\misfit)$ represents the current reconstructed 
wave speed at iteration $k$ using the misfit functional $\misfit$,
and $c^\dagger$ stands for the true wave speeds, displayed in 
\cref{fig:frgwi:true_models_cp,fig:frgwi:true_models_cs}.
Here, to compute the $L^2$-norm $\Vert \cdot \Vert$,
the two-dimensional models are represented on a Cartesian grid with 
$\num{2251}\times\num{901}=\num{2028151}$ coefficients.

\begin{figure}[ht!] \centering
  \renewcommand{\plotwidth} {6.10cm}
  \renewcommand{\plotheight}{3cm}
  \pgfmathsetmacro{\xmin}   {1}
  \pgfmathsetmacro{\xmax}   {60}
  \renewcommand{\dataA}{Jl2}
  \renewcommand{\dataB}{JFRgWI}
  \renewcommand{\datafile}{misfit_2hz.txt}
  \renewcommand{\mysubcaption}{\textbf{a)}}
  \pgfmathsetmacro{\capx}  {0}  \pgfmathsetmacro{\capy}  {0.35}
  \subfloat{\label{fig:misfit-comparison:misfit} 
\begin{tikzpicture}
\begin{axis}[
             enlargelimits=false, 
             ylabel={normalized misfit},
             xlabel={iterations at $2$ \si{Hz}},
             enlarge y limits=false,
             enlarge x limits=false,
             xmin=\xmin,xmax=\xmax,
             xtick={10,30,50},
             yminorticks=true,
             ytick={0.45,1},
             height=\plotheight,width=\plotwidth,
             label style={font=\scriptsize},
             tick label style={font=\scriptsize},
             legend style={font=\scriptsize\selectfont},
             scale only axis,
             ylabel style = {yshift =-0.50cm, xshift=1mm},
             legend pos={south west}, 
             clip mode=individual,
             ]  

     \pgfmathsetmacro{\scale}{6.535062}
     \addplot[color=\myblue,line width=1]
              table[x = iterations,y expr=1/\scale*\thisrow{\dataA},
                    ]
              {\datafile}; \addlegendentry{$\misfit_0$}    
     \pgfmathsetmacro{\scale}{4.719776}
     \addplot[color=\myred,line width=1]
              table[x = iterations,y expr=\thisrow{\dataB}/\scale]
              {\datafile}; \addlegendentry{$\misfit$}    

\node[text width=2cm,anchor=north west,black] at (\capx,\capy) {\mysubcaption};
\end{axis}
\end{tikzpicture}}
  \renewcommand{\dataA}{err_l2_cp}
  \renewcommand{\dataB}{err_frgwi_cp}
  \renewcommand{\dataC}{err_l2_cs}
  \renewcommand{\dataD}{err_frgwi_cs}
  \renewcommand{\datafile}{relative_error_models_2hz.txt}
  \renewcommand{\mysubcaption}{\textbf{b)}}
  \pgfmathsetmacro{\capx}  {0}  \pgfmathsetmacro{\capy}  {127.5}
  \subfloat{\label{fig:misfit-comparison:models} 
\begin{tikzpicture}
\begin{axis}[
             enlargelimits=false, 
             ylabel={{\normalsize{$\mathfrak{e}$}}},
             xlabel={iteration $k$ at $2$ \si{Hz}},
             enlarge y limits=false,
             enlarge x limits=false,
             xmin=\xmin,xmax=\xmax,
             xtick={10,30,50},
             ytick={138,129},
             yminorticks=true,
             height=\plotheight,width=\plotwidth,
             label style={font=\scriptsize},
             tick label style={font=\scriptsize},
             legend style={font=\scriptsize\selectfont},
             scale only axis,
             ylabel style = {yshift =-0.50cm, xshift=1mm},
             legend pos={south west}, 
             clip mode=individual,
             legend columns=2,
             ]  

     \pgfmathsetmacro{\scale}{1}
     \addplot[color=\myblue,line width=1]
              table[x = iterations,y expr=\thisrow{\dataA}*\scale,
                    ]
              {\datafile}; \addlegendentry{$\mathfrak{e}_k(c_p,\misfit_0)$}    

     \addplot[color=\myred,line width=1]
              table[x = iterations,y expr=\thisrow{\dataB}]
              {\datafile}; \addlegendentry{$\mathfrak{e}_k(c_p,\misfit)$}    

     \addplot[color=\myblue,line width=1,dashed]
              table[x = iterations,y expr=\thisrow{\dataC}]
              {\datafile}; \addlegendentry{$\mathfrak{e}_k(c_s,\misfit_0)$}

     \addplot[color=\myred,line width=1,dashed]
              table[x = iterations,y expr=\thisrow{\dataD}]
              {\datafile}; \addlegendentry{$\mathfrak{e}_k(c_s,\misfit)$}

\node[text width=2cm,anchor=north west,black] at (\capx,\capy) {\mysubcaption};
\end{axis}
\end{tikzpicture}}
  \caption{
  Comparison of performance in the minimization of 
  the misfit functionals $\misfit$ and $\misfit_0$ 
  in \cref{eq:misfit_strain,eq:misfit_raw}, respectively.
  The minimizations use the exploration data at $2$ \si{\Hz},
  starting from the initial models built with the FRgWI method 
  and the low-frequency regional earthquake
  data shown in \cref{fig:frgwi:tele_2hz_cp,fig:frgwi:tele_2hz_cs}.
  In order to use $\misfit_0$, the computational sources
  must coincide with the observational ones; the iterative 
  minimization relies on the non-linear conjugate gradient 
  method \citep{Nocedal2006}.
  \protect\subref{fig:misfit-comparison:misfit} 
    Evolution of the normalized misfit functionals (i.e., 
    scaled with its value at the first iteration) and    
  \protect\subref{fig:misfit-comparison:models} 
    relative error for the P- and S-wave speeds (\cref{eq:model-error}),
    for the iterations at $2$ \si{\Hz} frequency.}
  \label{fig:misfit-comparison}
\end{figure}

FRgWI improves drastically the convergence rate of the 
iterative minimization compared to the standard misfit 
function $\misfit_0$. We see that $\misfit_0$ stagnates 
rapidly in \cref{fig:misfit-comparison:misfit}, after 
about \num{15} iterations while in the FRgWI method, 
$\misfit$ keeps decreasing even after 60 iterations.
The improvement given by the FRgWI method is confirmed 
in terms of model error in \cref{fig:misfit-comparison:models}, 
which keeps decreasing for both the P- and S-wave speed models. 
We further note how the P-wave speed model keeps 
improving compared to the S one.
On the contrary, the relative model error associated 
to $\misfit_0$ stagnates after a few iterations, 
similar to the cost function in \cref{fig:misfit-comparison:misfit}.

In \cref{fig:misfit-comparison:l2}, we show the reconstructed 
P- and S-wave speeds obtained by minimizing $\misfit_0$ 
in \cref{eq:misfit_raw}, using the exploration data set with 
frequencies from $2$ \si{\Hz} to $10$ \si{\Hz}.
The inversion procedure starts from the initial models built
with the FRgWI method and the regional earthquake
data displayed in
\cref{fig:frgwi:tele_2hz_cp,fig:frgwi:tele_2hz_cs}.
Indeed, the stage one (inversion using the low-frequency regional earthquake
data) can only be performed with the FRgWI method, which does not
necessitate the knowledge (positions and functions) of the 
passive sources, hence we compare the minimization for 
the stage two only.

\begin{figure}[ht!] \centering
  \renewcommand{\modelfile}{cp-l2_10hz}  
  \renewcommand{\mysubcaption}{\textbf{a)}}
  \subfloat{\label{fig:misfit-comparison:l2-cp}
            \begin{tikzpicture}

\pgfmathsetmacro{\cmin}   {2.50}
\pgfmathsetmacro{\cmax}   {5.50}

\begin{axis}[%
  width =\modelRwidth,
  height=\modelRheight,
  axis on top, separate axis lines,
  xmin=\xminloc, xmax=\xmaxloc, xlabel={$x$   (\si{\km})},
  ymin=\zminloc, ymax=\zmaxloc, ylabel={depth (\si{\km})}, y dir=reverse,
  xlabel style={yshift=2mm},
  ylabel style={yshift=-2mm},
  y dir=reverse,ytick={0,2,4},
  colormap/jet,colorbar,
  colorbar style={title={$c_p$ (\si{\km\per\second})},
  title style={yshift=-2mm, xshift=-10mm},
  xshift=-.2cm, width=.2cm,
  },point meta min=\cmin,point meta max=\cmax
]
\addplot [forget plot] graphics [xmin=\xminloc,xmax=\xmaxloc,ymin=\zminloc,ymax=\zmaxloc] 
                                {{\modelfile}.png};

\end{axis}
\node[text width=2cm,anchor=north west,black] at (\capx,\capRy) {\mysubcaption};
\end{tikzpicture}%
} \hspace*{.5em}
  \renewcommand{\modelfile}{cs-l2_10hz} 
  \renewcommand{\mysubcaption}{\textbf{b)}}
  \subfloat{\label{fig:misfit-comparison:l2-cs}
            \begin{tikzpicture}

\pgfmathsetmacro{\cmin}   {1.70}
\pgfmathsetmacro{\cmax}   {3.20}

\begin{axis}[%
  width =\modelRwidth,
  height=\modelRheight,
  axis on top, separate axis lines,
  xmin=\xminloc, xmax=\xmaxloc, xlabel={$x$   (\si{\km})},
  ymin=\zminloc, ymax=\zmaxloc, 
  y dir=reverse, 
  xlabel style={yshift=2mm},
  y dir=reverse,ytick={0,2,4},
  colormap/jet,colorbar,
  colorbar style={title={$c_s$ (\si{\km\per\second})},
  title style={yshift=-2mm, xshift=-10mm},
  xshift=-.2cm, width=.2cm,ytick={2,3},
  },point meta min=\cmin,point meta max=\cmax
]
\addplot [forget plot] graphics [xmin=\xminloc,xmax=\xmaxloc,ymin=\zminloc,ymax=\zmaxloc] 
                                {{\modelfile}.png};

\end{axis}
\node[text width=2cm,anchor=north west,black] at (\capx,\capRy) {\mysubcaption};
\end{tikzpicture}%
}
  \caption{
           The reconstructed \protect\subref{fig:misfit-comparison:l2-cp}
           P-wave speed and \protect\subref{fig:misfit-comparison:l2-cs} 
           S-wave speed models obtained from the minimization of $\misfit_0$ 
           in \cref{eq:misfit_raw}. 
           The inversion uses the exploration data from $2$ \si{\Hz} 
           to $10$ \si{\Hz}, starting from the models built with the 
           FRgWI method and the low-frequency regional earthquake data in
           \cref{fig:frgwi:tele_2hz_cp,fig:frgwi:tele_2hz_cs}.
           The reconstructions obtained with the FRgWI method
           and minimization of $\misfit$ in \cref{eq:misfit_strain} 
           are displayed in \cref{fig:frgwi:explo_cp,fig:frgwi:explo_cs}
           for comparison.
          }
  \label{fig:misfit-comparison:l2}
\end{figure}

The rapid stagnation of the misfit function and model errors 
in \cref{fig:misfit-comparison} is confirmed by the pictures 
of the properties reconstructed (\cref{fig:misfit-comparison:l2-cp,fig:misfit-comparison:l2-cs})
from the minimization of $\misfit_0$ in \cref{eq:misfit_raw} 
using frequencies from $2$ \si{\Hz} to $10$ \si{\Hz}.
We see that the iterative minimization algorithm barely modifies 
the initial models given in \cref{fig:frgwi:tele_2hz_cp,fig:frgwi:tele_2hz_cs}.
These reconstructed models are particularly inaccurate compared 
to the reconstructions obtained from the minimization of $\misfit$ 
in \cref{eq:misfit_strain} which are pictured in \cref{fig:frgwi:explo_cp,fig:frgwi:explo_cs}.

In the same configuration, that is, when the computational
sources are selected to coincide with the observational ones
and when we start from the same initial guess, we see that 
the FRgWI method behaves better than the least-squares criterion. 
In addition, the FRgWI allows for arbitrary computational sources, 
thus offering more flexibility: for instance the stage one of the 
inversion using the regional earthquake data cannot be performed with 
the least-squares criterion of \cref{eq:misfit_raw}.

\section{Discussion}

The main feature of FRgWI is that the sources that generate the 
measurements do \emph{not} have to correspond with the sources that 
are used for the simulations and the misfit function then tests each 
of the measured data with every simulation.
In this work, we have experimented the use of data from 
events (observational sources) located outside the 
computational domain without requiring any pre-processing.
With FRgWI, we do \emph{not} need to know the 
precise origin of the (passive) source
and we do \emph{not} have to produce simulations on the large 
domain where the event has occurred.
Contrary to conventional teleseismic inversion 
which needs distant earthquake to work with plane wave 
illumination, FRgWI allows for any type of passive 
sources, hence enabling the use of data from local events.
The use of data from regional (or teleseismic) events
compensates for the lack of low-frequency contents in the exploration data, 
and are used as a first step to build initial models.
As the method requires numerical simulations of the displacement
and stress, the HDG discretization is an appropriate 
candidate to solve numerically the forward problem, by 
working with the first-order system at a reduced computational cost.
We have implemented the method in the frequency domain, which can 
make it difficult to handle large media in three-dimensions, due 
to the memory requirement. Nonetheless, 
the method can similarly work in the time (or hybrid) domain, using 
a frequency continuation approach \citep{Bunks1995}.
The resulting cross-correlations in the misfit function should 
not lead to any additional computational difficulties 
and the time domain allows to account for larger domains.

We have provided a pilot study to illustrate the possibilities 
offered by the FRgWI method: it now needs to be 
applied on more practical settings.
While the formulation of the misfit functional in terms of 
the stress remains identical with anisotropy (\cref{eq:misfit_stress}), 
it implies that measurements of the complete strain tensor
have to be obtained.
The misfit functional is defined from the integral over the receiver 
surface such that, in the case of topography, the method can be applied provided 
the surface is Lipschitz.
The misfit can be evaluated via a quadrature rule, that is, weighting 
the contribution of the receivers depending on their positions.
This is the subject of ongoing research.

Another possible extension of the FRgWI method is 
by including attenuation, which
is relatively straightforward in the frequency-domain: its
incorporation maintains the constitutive law in \cref{eq:hookes}
with a complex-valued elasticity tensor \citep{Bland1960,Carcione2007}.
However, in the time-domain, the constitutive law for a medium 
with attenuation incorporates time derivatives of the strain 
and/or stress \citep{Carcione2007}, and the reciprocity 
formula would have to be derived accordingly.
While the misfit functional allows to work with observational sources
at arbitrary positions, as we have illustrated, it also allows for the 
arbitrary positions and functions of the computational ones.
Consequently, FRgWI a good candidate for shot-stacking (source
summation), as highlighted by \cite{Faucher2019FRgWI} in the acoustic case.
In addition, while we have used computational sources near the surface only,
using alternative positions should be investigated, as well as other
source types and geometries (e.g., using plane-waves). 
Namely, the experiment we have carried out is the first step to
analyze the potential of the method.

\section{Conclusion}

We implement a new misfit functional for elastic 
reconstruction of Earth parameters,
and have shown a preliminary experiment which 
illustrates its flexibility regarding the 
acquisition setups.
The underlying FRgWI method opens up the 
perspective of considering 
passive-source data for exploration, in order to 
recover the low-wavenumbers subsurface models.
These low-frequencies, missing in the exploration data, 
are crucial to build smooth background models 
and to mitigate cycle-skipping issues during inversion.

Applications of the FRgWI method directly relates 
to the ability of the new acquisition techniques
with fiber optic cable in DAS to obtain the strain. 
Nonetheless, it is still difficult to 
rely on the availability of such data for practical 
applications. 
For instance, the precise orientation of the fiber
can be hard to control. Our study remains conceptual 
at this stage
and the accuracy of the required DAS technology and  
the available bandwidth of such measurements must be addressed.
Then, the implementation of the method from the strain also 
requires the values of the medium parameters (except the density)
along the cable. 
Future applications should follow the steps:
(1) reconstruction of the very near-surface elastic 
    parameters using the (surface) exploration data, 
(2) minimization of the sub-surface reciprocity functional 
    using the regional earthquake data to recover the 
    low wavenumbers 
    of the models and 
(3) reconstruction of the high wavenumbers 
    with the exploration data.
In this work, we carried out a pilot computational experiment 
to study the capability of our approach and to illustrate how 
to exploit DAS. 


\section*{Acknowledgments}

The authors also thank Jonathan Ajo-Franklin for 
interesting discussions on Distributed Acoustic 
Sensing. 
The research of FF is supported by the Austrian Science 
Fund (FWF) under the Lise Meitner fellowship M 2791-N.
MVdH was supported by the Simons Foundation under the MATH+X 
program, the National Science Foundation under grant DMS-1815143, 
and the corporate members of the Geo-Mathematical Imaging 
Group at Rice University, USA. 
OS is supported by the FWF, with SFB F68, project F6807-N36 
(Tomography with Uncertainties).

The code used for the experiments, \texttt{hawen}, is 
developed by the first author and available 
at \texttt{https://ffaucher.gitlab.io/hawen-website/}.
The scripts and data-set to reproduce the results are 
archived at \texttt{http://phaidra.univie.ac.at/o:1097637}.

\bibliographystyle{apalike}
\bibliography{bibliography}

\end{document}